\documentclass[9pt, conference]{IEEEtran}

\usepackage[pagebackref=false,breaklinks=true,colorlinks,linkcolor=red,citecolor=citecolor,urlcolor=blue,bookmarks=false]{hyperref}
\def\BibTeX{{\rm B\kern-.05em{\sc i\kern-.025em b}\kern-.08em
    T\kern-.1667em\lower.7ex\hbox{E}\kern-.125emX}}

\usepackage{algorithm}
\usepackage{graphicx}
\usepackage{subfig}
\usepackage{threeparttable}
\usepackage{color}
\usepackage{colortbl}
\usepackage{multirow}
\usepackage{amsmath}
\usepackage{amsfonts}
\usepackage{comment}
\usepackage{bm}
\usepackage{etoolbox}                                      
\usepackage{bbm}
\usepackage{url}
\usepackage{nth}
\usepackage{balance}
\usepackage{courier}                                       
\usepackage{booktabs}                                      
\usepackage{listings}                                      
\usepackage[]{algpseudocode}
\usepackage{upgreek}
\usepackage{xspace}
\usepackage[noadjust]{cite}

\algdef{SE}[DOWHILE]{Do}{doWhile}{\algorithmicdo}[1]{\algorithmicwhile\ #1}%

\newcommand{\name}{\texttt{ELight}\xspace}
\graphicspath{{./figs/}}
\definecolor{citecolor}{RGB}{34,139,34}
\begin{document}
\bstctlcite{IEEEexample:BSTcontrol} 

\pagestyle{plain} 

\title{ELight: Enabling Efficient Photonic In-Memory Neurocomputing with Life Enhancement}

\author
{
Hanqing Zhu\textsuperscript{$\star$},
Jiaqi Gu,
Chenghao Feng,
Mingjie Liu,
Zixuan Jiang,
Ray T. Chen,
and
David Z. Pan\textsuperscript{$\dagger$}
\\
ECE Department, The University of Texas at Austin, Austin, TX, USA \\
\textsuperscript{$\star$}hqzhu@utexas.edu; \textsuperscript{$\dagger$}dpan@ece.utexas.edu
}

\maketitle

\begin{abstract}
\label{abstract}
With the recent advances in optical phase change material (PCM), photonic in-memory neurocomputing has demonstrated its superiority in optical neural network (ONN) designs with near-zero static power consumption, time-of-light latency, and compact footprint.
However, photonic tensor cores require massive hardware reuse to implement large matrix multiplication due to the limited single-core scale.
The resultant large number of PCM writes leads to serious dynamic power and overwhelms the fragile PCM with limited write endurance. 
In this work, we propose a synergistic optimization framework, \name, to minimize the overall write efforts for efficient and reliable optical in-memory neurocomputing.
We first propose write-aware training to encourage the similarity among weight blocks, and combine it with a post-training optimization method to reduce programming efforts by eliminating redundant writes. 
Experiments show that \name can achieve over $20 \times$ reduction in the total number of writes and dynamic power with comparable accuracy.
With our \name, photonic in-memory neurocomputing will step forward towards viable applications in machine learning with preserved accuracy, order-of-magnitude longer lifetime, and lower programming energy.
\end{abstract}

\section{Introduction}
\label{sec:Introduction}
Optical neural networks (ONNs)~\cite{NP_NATURE2017_Shen, NP_ASPDAC2019_Zhao, NP_ASPDAC2020_Gu, NP_DATE2019_Liu, NP_PIEEE2020_Cheng, NP_DATE2020_Zokaee,NP_Nature2020_Wetzstein,NP_DATE2021_Gu, NP_DATE2021_Gu2, NP_NaturePhotonics2021_Shastri, NP_TCAD2020_Gu} are widely studied as a promising neurocomputing paradigm  with ultra-high speed, ultra-low energy cost, and high bandwidth to satisfy computation demands of machine learning applications.
Recent work~\cite{NP_APR2020_Miscuglio, NP_Nature2020_Feldmann} demonstrates that phase change material (PCM) can be used to build photonic tensor cores (PTCs) for optical in-memory matrix multiplication.
PCM cells are programmable with non-volatile states as a weight encoding mechanism.
By shining light through waveguides integrated with configured PCMs, light-matter interactions will change the amount of light transmission passively, thus achieving in-memory multiplication with near-zero static power.
Moreover, the broadband transmission of PCMs enables massive parallelism with wavelength-division multiplexing (WDM).
Holding above superiority,
PCM-based PTCs open a new pathway towards efficient in-memory neurocomputing via photons.

\begin{figure}
    \centering
    \vspace{-5pt}
    \subfloat[]{\includegraphics[width=0.225\textwidth]{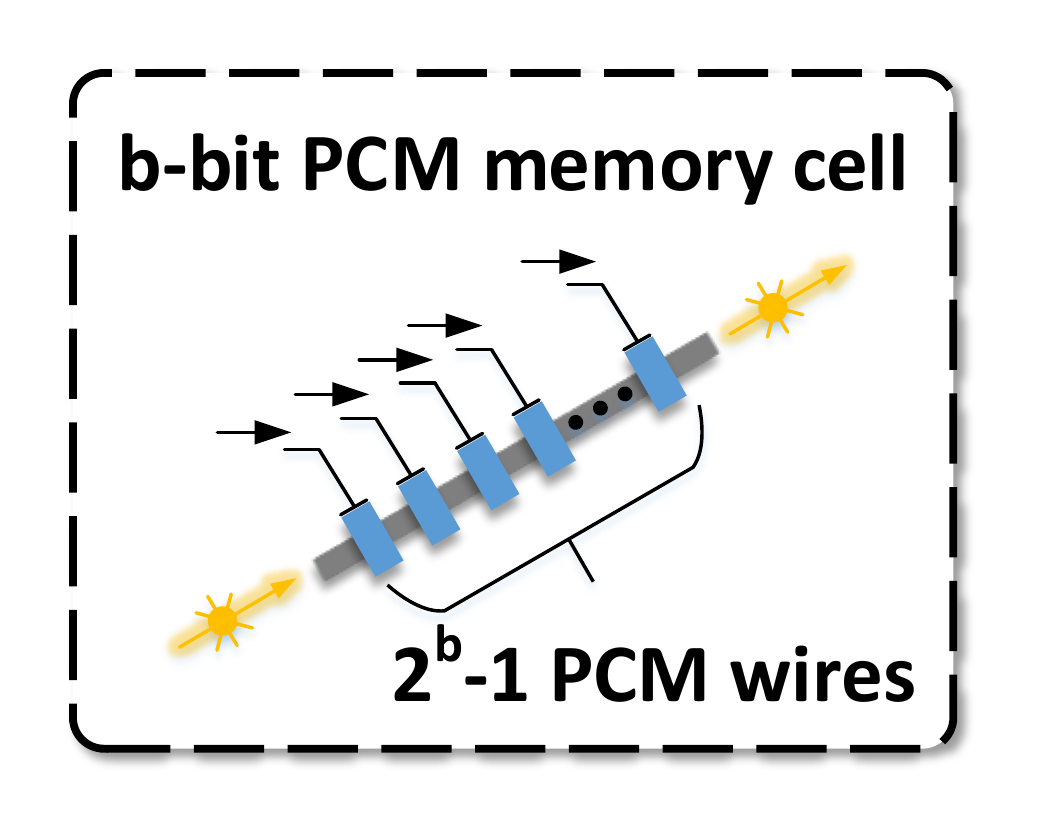}
    \label{fig:PhotonicMemoryPCM}
    }
    \subfloat[]{\includegraphics[width=0.225\textwidth]{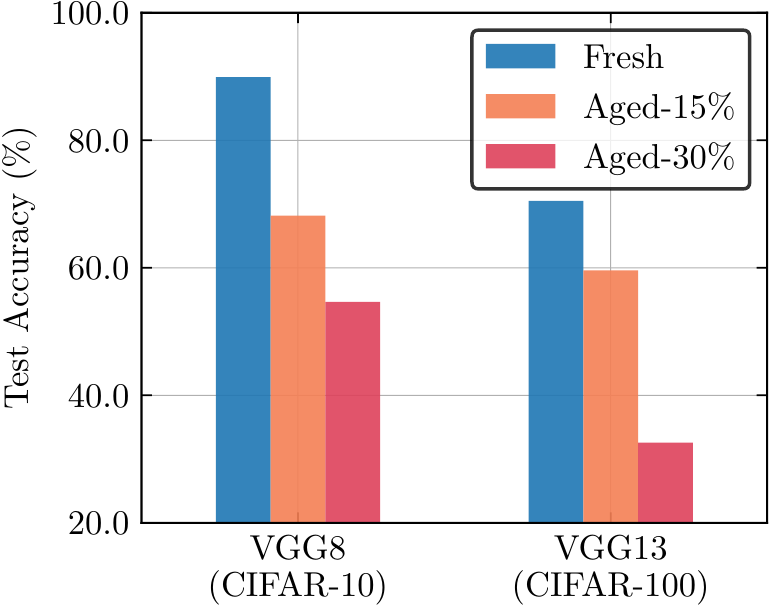}
    \label{fig:stuckFAccDrop}
    }
    \vspace{-5pt}
    \caption{~\small 
    (a) Multi-level photonic memory cell based on PCM wires.
    (b) Accuracy comparison of models with/without a fixed ratio of aged cells.}
    \label{fig:Motivition}
    \vspace{-10pt}
\end{figure}

However, PCM-based photonic in-memory neurocomputing still encounters practical barriers before truly viable for efficient inference acceleration.
Firstly, current PCM cell designs support only limited bit-width data imprint.
In \cite{NP_APR2020_Miscuglio}, a reasonable implementation of $b$-bit PCM cell with low programming complexity is proposed for PTCs, with demonstrated $4$$\sim$$5$-bit storage.
It is equivalently realized by patterning 2$^b$$-$$1$ identical PCM wires on the same waveguide, as shown in Fig.~\ref{fig:PhotonicMemoryPCM}.
Each wire represents binary phase states by completely \emph{crystallization} or \emph{amorphization}, individually switched via \emph{electrothermal heating}, 
thus allowing total $2^b$ nonlinear transmission levels.
Hence, a specialized quantization strategy that suits the unique transmission level distribution is in high demand to preserve the ONN accuracy on the above low-precision PCM-based PTCs.

Besides limited bit-width, the potential frequent weight reprogramming during inference also raises critical issues in PCM-based photonic in-memory computing.
Massive reuse of PTCs is required due to the limited scale of PTC compared to the weight matrix, e.g., a $64\times64$ PTC is already quite large.
The resultant massive weight updates in PTCs potentially threaten PCM wires at a high risk of over-utilization, given limited write endurance of PCM, ranging from $10^6$ to $10^8$~\cite{NP_APL2021_Zhang}.
Once PCM wires are aged and lose reprogrammability, the implementable transmission range will degrade, and the physical value will deviate from the desired value, leading to severe accuracy drops, shown in Fig.~\ref{fig:stuckFAccDrop}.
Moreover, the massive reprogramming has a non-trivial dynamic energy cost, which dilutes the energy efficiency benefits from PCM.
The above issues are related to two key metrics:
(1) The average utilization of PCM wires indicated by the total number of write operations (\emph{\# total writes});
(2) The maximum number (\emph{\# max writes}) of write operations over a single PCM cell in one PTC.

In this work, we propose a synergistic aging-aware optimization framework \name to tackle the issues. 
Based on an augmented redundant write elimination strategy, we devote to trimming down \emph{redundant} writes on PCM wires in photonic memories during weight updates so as to reduce \emph{\# total writes} and \emph{\# max writes}.
Aware of the block pattern of weight reloading,
we first propose a write-aware training method to orchestrate the higher similarity among weight blocks to increase the eliminable redundant writes.
Then post-training optimization is applied to reduce the number of writes further.
The main contributions of this paper are listed as follows.

\begin{itemize}
    \item \textbf{Distribution-Aware Quantization} scheme is introduced to reduce weight encoding errors on PCM cells with the awareness of modeled transmissivity distribution.
    \item \textbf{Write-Aware Training} is proposed to boost block-wise weight similarity and reduce redundant writes during weight updates with negligible effect on accuracy.
    \item \textbf{Post-Training Optimization} is employed to further cut down redundant writes via column-based reordering, without changing the model output.
    \item To the best of our knowledge, this is the \emph{first work} that handles the aging and energy efficiency issue of PCM-based photonic neural engines.
    Our \name achieves over $20\times$ reduction in the total number of re-programming operations and dynamic energy cost during inference, enabling long-life and efficient photonic in-memory neurocomputing.
\end{itemize}

\section{Preliminaries}
\label{sec:Background}
In this section, we introduce the basics of phase change material, the architecture of photonic tensor core, and current barriers in the practical deployment of PCM-based ONNs.

\begin{figure}[]
    \centering
    \includegraphics[width=0.49\textwidth]{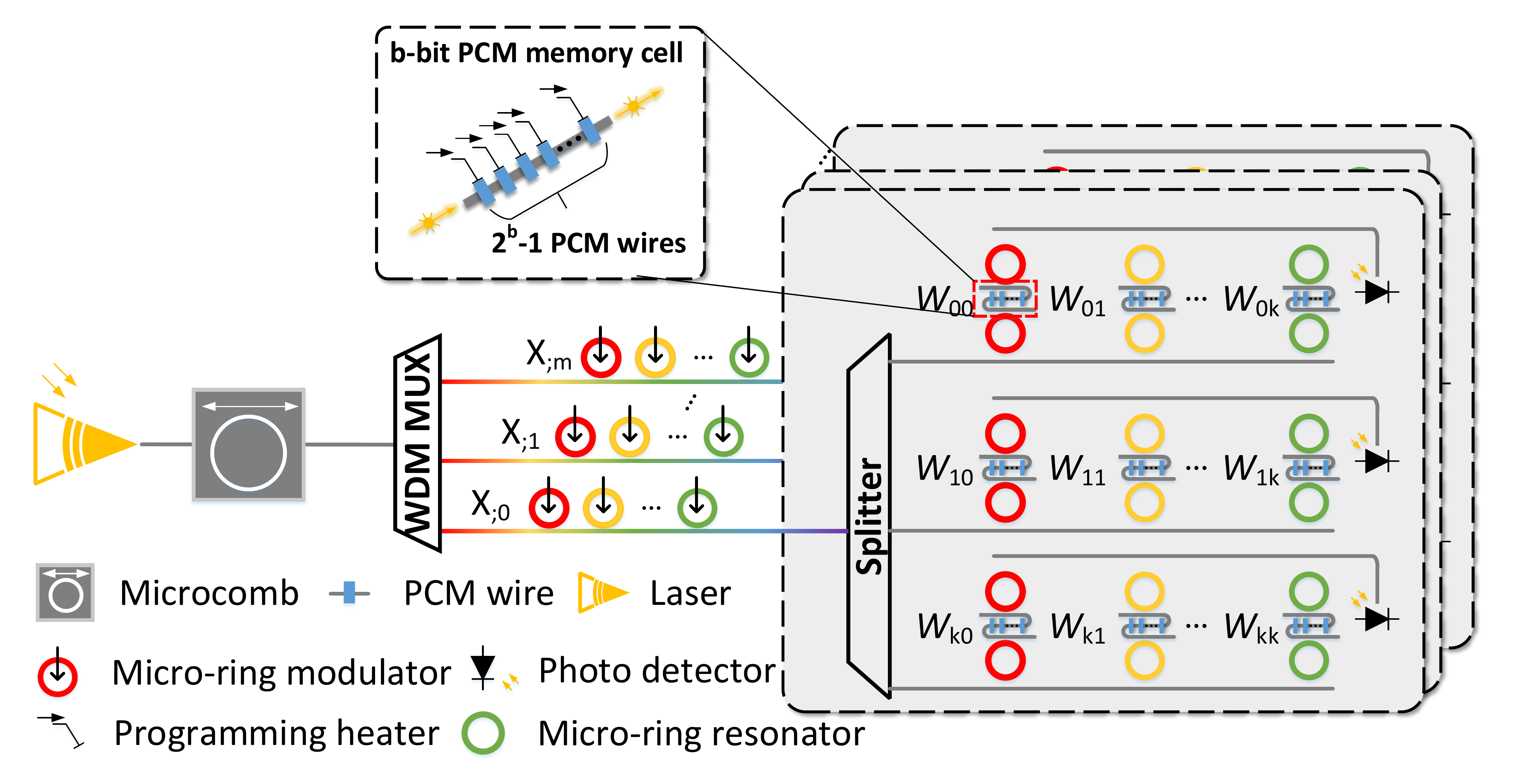}
        \caption{The architecture of photonic tensor engines based on PCM.}
    \label{fig:pcmTensorCore}
\end{figure}

\subsection{Basics of Phase Change Material (PCM)}
As a promising memristive device, phase change material (PCM) draws particular attraction in photonic in-memory computing.
PCM can cause a pronounced change in the optical property and switch between low-light-transmission crystalline (c) state, representing a logical `0', and high-light-transmission amorphous (a) state, representing a logical `1'.
The programmable non-volatile states endow PCM with a potential to demonstrate ultra-fast in-memory multiplication \cite{NP_APR2020_Miscuglio, NP_Nature2020_Feldmann}.
By shining light through the waveguide with PCM cells on top, the transmitted optical power can be modulated as $P_{out} = t \cdot P_{in}$ with $t$ being the transmission factor inscribed in PCM states, where the scalar multiplication is implemented.
Yet, the number of realized distinct transmission levels in current PCM cell designs is rather limited for high-precision data storage.
In~\cite{NP_APR2020_Miscuglio}, an implementation of $b$-bit PCM cells with $2^b$ transmission levels is proposed for in-memory photonic analog computing by patterning 2$^b$$-$$1$ binary PCM wires on one waveguide.
Considering the practicality of binary PCM devices and low programming complexity, we will focus on this PCM cell design~\cite{NP_APR2020_Miscuglio} in the following discussion.
However, PCM suffers from limited endurance with a maximum of $10^6$$\sim$$10^8$ total reprogramming times~\cite{NP_APL2021_Zhang}.
Frequent writing operations will over-utilize PCM, shorten its lifetime, and reduce reliability due to the loss of reprogrammability.

\subsection{Architecture of Photonic Tensor Core}
Recent work~\cite{NP_APR2020_Miscuglio, NP_Nature2020_Feldmann} demonstrates the implementation of photonic tensor cores for optical in-memory matrix multiplication, i.e, $Y=WX+b$.
Both utilize the photonic PCM arrays' storage and light interaction capability, with $W$ being encoded in the PCM states.

Consider the weight block $W\in \mathbb{R}^{k\times k}$ and input matrix $X$ with $m$ columns.
Fig.~\ref{fig:pcmTensorCore} illustrates one architecture of PCM-based PTC~\cite{NP_APR2020_Miscuglio}, where matrix-matrix multiplication (MM) is achieved by duplicating PCM array $m$ times to carry out multiple matrix-vector multiplication (MVM).
Starting with an input laser and an on-chip frequency comb to generate multiple wavelengths ($\lambda_{0}, \lambda_{1}, \cdots$), a WDM multiplexer is used to distribute the light into $m$ rows evenly. 
For each row, a series of narrowband micro-ring modulators are used to imprint one column of the input matrix $X$ in the
power of the optical signals.
Then, a PCM array with $k$ rows and $k$ columns can be used to encode the weight block and achieve MVM between $W$ and one column of $X$.
Concretely, in the $i$-th rail, the inputs are filtered by the on-resonance micro-rings and weighted via light-matter interaction with the PCM.
At the end of the drop port, photo-detectors are used to accumulate the intensity of the WDM optical signals, i.e., the weighted inputs, and output the desired result $Y_{im} = \sum_j^k W_{ij}X_{jm}$. 

In Fig.~\ref{fig:pcmTensorCore}, the $b$-bit PCM cell is equivalently realized by patterning $2^b$$-$$1$ binary PCM wires on the same waveguide to enable $2^b$ transmission levels.
To store the desired value, the photonic memory cell is programmed by selectively switching the phase of the wires between crystalline and amorphous states.
For example, for a 4-bit photonic memory with fifteen PCM wires, '1100' is demonstrated by randomly programming twelve wires to the high-light-transmission amorphous state while setting others to the crystalline state.
Note that the electrothermal a-c and c-a transition of PCM wires is achieved by sending electrical pulses to individual thermal heaters in parallel.

\subsection{Barriers in Practical Deployment of PCM-based ONNs}
As a promising platform for machine learning (ML), photonic in-memory neurocomputing still encounters practical challenges.
First, the limited programming resolution of  PCM-based PTCs calls for a specialized quantization scheme.
Besides, the limited scale of PTC cannot promise the one-shot realization of large matrix multiplication.
A $64\times 64$ PTC is already quite large due to area cost and light loss~\cite{NP_Nature2020_Feldmann}.
Hence, massive reuse of PTCs is required during inference, leading to frequent weight updates on PCM arrays.
Consequently, with limited endurance, PCM wires are at high risk of over-utilization, thus shortening the lifetime of photonic tensor cores.
Moreover, a massive number of write operations requires significant dynamic power, which might raise concerns about the energy superiority claimed for PCM.

In this paper, to tackle the above issues, we trace and optimize two key metrics.
The total number of write operations of PCM wires (\emph{\# total writes}) reflects the averaged degree of utilization of PCM wires and the dynamic energy cost.
The maximum number of wire write operations of a single photonic cell (\emph{\# max writes}) can represent the status of the most over-utilized memory cell, determining the lifetime of PTC.
Unlike previous works~\cite{NP_DAC2018_cai, NP_ASPDAC2019_Xia, NP_ICCD2019_wen, NP_TCAD2018_Meng}, which focus on optimizing frequent weight updates when the training is performed on emerging neuromorphic computing systems such as ReRAM, 
in this work,
we focus on the potential frequent weight reprogramming during the inference process, which plagues photonic in-memory neurocomputing as a curse of the limited scale of PTCs.

\section{Proposed Distribution-Aware Quantization Scheme}
\label{sec:adaption}

In this section, we give out a dedicated \emph{distribution-aware quantization scheme} based on the analysis of transmission level distribution of $b$-bit PCM memory cell, to reduce weight encoding errors.

\subsection{Transmission Model of Multi-Level PCM Memory Cell}
The light transmissivity of a $b$-bit photonic memory cell is determined by the phase states of $2^b$$-$$1$ PCM wires. 
Assuming that level $i$ refers to the condition that $i$ wires are set to $c$ state while the others are set to $a$ state, 
its extinction ratio (ER) is computed as the ratio of the transmitted optical power in level $i$ and level $0$, i.e., $10\log_{10}(\frac{P_i}{P_0})$.
As demonstrated in~\cite{NP_APR2020_Miscuglio}, the ER uniformly increases as a function of $i$ with a step $\Delta e$.
Given that the transmitted optical power can be written as the product of the transmission factor $t_i$ and input light power, ER can be further expressed as
\begin{equation}
    \small
    10\log_{10}(\frac{t_i \cdot P_{in}}{t_0 \cdot P_{in}}) = 10\log_{10}(\frac{t_0 \cdot P_{in}}{t_0 \cdot P_{in}}) + i \Delta e.
\end{equation}
Thus, the $i$-th transmission level can be derived as 
\begin{equation}
    \small
    t_i = t_0\cdot 10^{\frac{1}{10}\Delta e \times i} = t_0 \times c^i, \, c = 10^{\frac{1}{10}\Delta e}.
\end{equation}
Here, $c\in(0,1)$ indicates the percentage of light power transmitted through one PCM wire.
The $0$-th transmission level $t_0$ corresponds to all wires being in $a$ state, which is approximately 1~\cite{NP_APR2020_Miscuglio}.

Hence, we can finally formulate the distribution of $2^b$ transmission levels of $b$-bit PCM memory cell as an \emph{exponential model},
\begin{equation}
\small
    t_{i} = c^i, \ i= 0, 1, \dots, 2^b-1.
\end{equation}

\subsection{Augmented Base-$c$ Quantization}
An important question is how to effectively map full-precision weights $w$ to the exponential transmission levels of PCM photonic memory with low quantization error.
Given the exponential transmissivity distribution, normal uniform quantization fails to fit it well with severe encoding error. 
Hence, a dedicated quantizer $q(w,b)$, where $b$ is the bit-width, is required to minimize the quantization error,
\begin{equation}
    \small
    \min~\lVert \hat{w} - w \rVert^2_{2}, \quad \text{s.t.~} \ \hat{w}= q(w, b) \in Q_{b},
    \label{eq:quantization_error}
\end{equation}
where $Q_{b}$ denotes a set of quantization levels, i.e., the transmission levels of a $b$-bit photonic memory cell.

However, as PCM can only demonstrate positive light transmission, to support full-range weight, we store the positive and negative values of weight matrix $W$ in the positive PTC and the negative PTC, respectively.
Then, the differential photo-detection module will generate balanced output.
With the simple but effective \emph{weight extension strategy}, each scalar weight $w$ in $W$ is expressed as 
\begin{equation}
\small
    w= w_{pos} - w_{neg}.
\label{eq:w_ext}
\end{equation}
Here, $w_{pos}$ and $w_{neg}$ are physical values in positive and negative photonic memory cells, where one is selected based on the sign of $w$ to store the weight value and the other is set to the lowest light transmission level $\delta$.
In this way, the differential weight encoding in \eqref{eq:w_ext} augments the quantization codebook $Q_b$ as follows,
\begin{equation}
\small
    Q_b = \{c^{2^b-1}-\delta, \pm( c^{2^b-2}-\delta), \dots, \pm (c^{0}-\delta)\},  \, \delta=c^{2^{b}-1},
\end{equation}
where the number of implementable quantization levels in $Q_b$ is almost doubled.
This attributes our quantization with a higher model expressivity, especially under low-bit quantization. 

With the augmented quantization codebook, for $w$ within $[-1, 1]$, an augmented base-c quantizer is hence proposed to optimize \eqref{eq:quantization_error} as
\begin{equation}
\small
    w_q = q(w, b) =  \frac{\texttt{sign}(w)}{s}\cdot( c^{\texttt{Clip}( \texttt{R}(\log_{c}(s|w|+\delta)), \ 0, \ 2^b-1)} - \delta),
\label{eq:quant_func}
\end{equation}
where the scaling factor $s=c^0 - c^{2^b-1}$ is used to transform quantization levels into $[-1, 1]$ and $\texttt{R}(\cdot)$ is a round function.
 $\texttt{Clip}(\cdot)$ is a clip function to limit the value within $[0, 2^b-1]$.
A quantization-aware training procedure \cite{NN_Arxiv2016_Zhou} is adopted with our augmented base-$c$ quantizer to train the PCM-based ONNs.
In the forward propagation, the weights $W$ are quantized as
\begin{equation}
\small
    w_q = q(\frac{\texttt{Tanh}(w)}{\max(\texttt{Tanh}(w))}, b), \ b>1.
\end{equation}
In the backward pass, we coarsen the whole b-bit quantization process
$q(w, b)$ as an entirety and estimate its gradient $g_q$ by \cite{NN_LectureSTE_Hinton}:
\begin{equation}
\small
    g_q = \frac{\partial \mathcal{L}}{\partial W} =  \frac{\partial \mathcal{L}}{\partial W_q} \frac{\partial W_q}{\partial W} = \frac{\partial \mathcal{L}}{\partial W_q}.
\end{equation}
Considering limited on-chip storage, a uniform quantizer in \cite{NN_ICLR2020_Stock} is used to discretize layer input.

\section{Proposed Aging-Aware Co-Optimization Framework}
\label{sec:Method}

In this section, we propose the aging-aware co-optimization framework, \name, to minimize both \emph{\# total writes} and \emph{\# max writes}.
We first illustrate the problem formulation and the adopted \emph{augmented redundant write elimination} (ARWE) strategy for wire-level writes in PCM cells.
Then we describe the proposed \emph{write-aware training} method to encourage the similarity among weight blocks.
At last, we propose a fine-grained \emph{column-based reordering} method further to cut down redundant writes as a post-training optimization strategy.

\subsection{Problem Formulation}
Due to the limited scale of a single PTC, we adopt blocking matrix multiplications to implement convolutional layers and fully-connected layers for practical considerations.
Specifically, an im2col algorithm~\cite{NN_MM14_Jia} converts the convolutions into general matrix multiplication (GEMM).
The weight matrix $W\in \mathbb{R}^{M\times N}$ is partitioned into $P\times Q$ sub-matrices, where each $k\times k$ block can be deployed onto one PTC.
Then the set $B$ of sub-matrices is assigned to a cluster $C$ of photonic tensor cores.
Since we have $M,N\gg k$, the number of sub-matrices is quite large.
For example, assuming that the size of PTC is $64\times 64$, $B$ contains $8\times 72$ blocks to implement the 5$^\mathrm{th}$ convolutional layer of VGG8.
Moreover, considering the limited number of on-chip PTCs 
, multiple sub-matrices need to be assigned to one PTC, leading to massive reuse of PTCs during inference.
Here, without loss of generality, we adopt an assignment strategy to assign sub-matrices to PTCs for the following discussion.
For one layer with $P\times Q$ blocks, 
a cluster of PTCs is dedicated for the MM,
where one PTC is assigned with a row of weight blocks.
$P$ PTCs carry out blocking MM in parallel with shared input.
We assume we write new block data into PTC after finishing all the block MMs on the current stored block to reduce reprogramming efforts.
It should be noted that our method can work with other assignment schemes.

\begin{figure}
    \centering
    \includegraphics[width=0.48\textwidth]{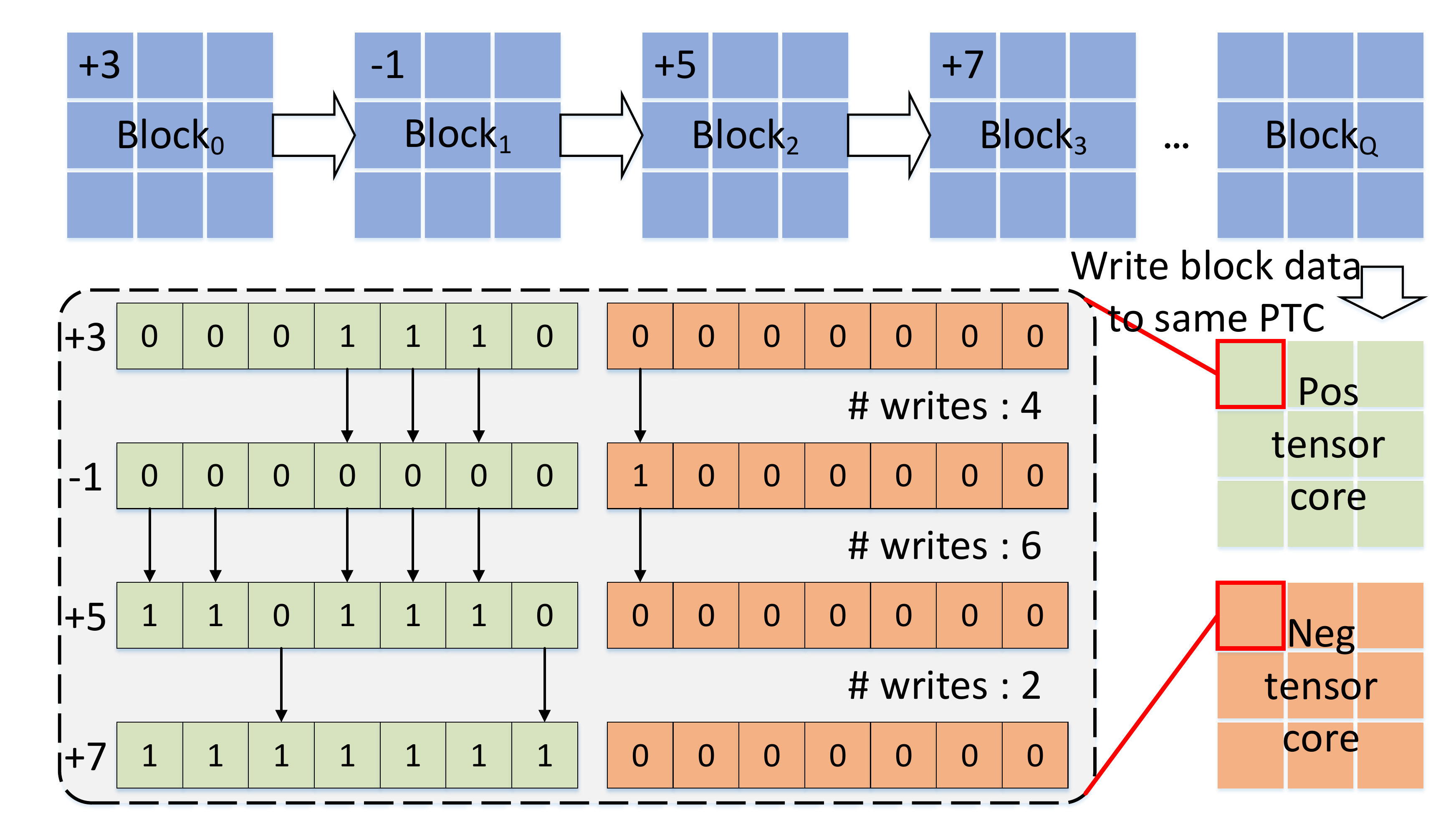}
    \caption{~\small The illustration of writing multiple weight blocks in the same PTC with redundant wire writes elimination.
    Each PCM wire within 3-bit memory is in a(1) or c(0) state to demonstrate quantization levels in -7$\sim$+7. 
    Wire-level redundant writes are excluded to reduce \emph{\# writes}.
    }
    \label{fig:write_ex}
\end{figure}

As data are represented by the binary-state PCM wires within photonic memories,
inspired by redundant write elimination (RWE) strategy~\cite{NP_ISCAs2007_Yang}, we propose an augmented redundant write elimination (ARWE) strategy,
where we eliminate the writes for identical values and eliminate identical wire-level writes.
Concretely, to demonstrate desired value, we preserve the current states of PCM wires at the largest extent by only perturbing the smallest number of wires.
For a clear illustration, Fig.~\ref{fig:write_ex} shows one example of writing a sequence of weight blocks into 3-bit PTCs.
One positive PTC and one negative PTC are used together to demonstrate full-range weights.
When programming transmission level +7 into the memory cell storing level +5, two $c$-state PCM wires in the positive PTC need to be programmed to $a$ state to achieve the smallest number of writes.
Note that the binary state of each PCM wire would be stored in memory using 1bit such that off-chip computers can reprogram based on ARWE strategy, without the need of detecting stored values.

Thus, considering write efforts in both positive and negative PTCs, the number of writes (WT) between two $b$-bit numbers $w^{'}$ and $w$ is computed as follows,
\begin{equation}
\small
\begin{aligned}
WT(w^{'}, w) &= |l^{+}(w^{'}) - l^{+}(w)| + |l^{-}(w^{'}) - l^{-}(w)|,
\end{aligned}
\label{eq:wt_eq}
\end{equation}
where $l^{+}$ and $l^{-}$ denote the transmission level in positive and negative PTC, respectively.
The absolute value of $l^{+}$ and $l^{-}$ also indicate the number of $a$-state wires out of 2$^b$$-$$1$ wires in positive and negative PTCs, respectively, 
which can be derived based on \eqref{eq:quant_func} as
{\small\begin{align}
    l^{+}(w) & =\left\{
\begin{aligned}
(2^b\!-\!1)\!-\!\texttt{Clip}( \texttt{R}(\log_{t}(s|w|+\delta)), 0, 2^b\!-\!1), \ w \geq 0 \\
0, \ w < 0
\end{aligned}
\right.  
\label{eq:level_pos}\\
l^{-}(w) & =\left\{
\begin{aligned}
0, \ w \geq 0 \\
\texttt{Clip}( \texttt{R}(\log_{t}(s|w|+\delta)), 0, 2^b\!-\!1)\!-\!(2^b\!-\!1), \ w < 0
\end{aligned}
\right..
\label{eq:level_neg}
\end{align}}
Then, we have transmission level of $w$ as $l(w) = l^{+}(w) - l^{-}(w)$, ranging from $-(2^b-1)$ to $(2^b-1)$.

Accordingly, to write new block $A^{'}$ of size $k\times k$ to photonic memories storing block $A$, the number of writes is computed as
\begin{equation}
\small
\begin{aligned}
WT(A^{'}, A)\!\!=\!\!\sum_{i}^{k}\sum_{j}^{k} (|l^{+}(a_{ij}^{'})\!-\!l^{+}(a_{ij})|+ |l^{-}(a_{ij}^{'})\!-\!l^{-}(a_{ij})|).
\end{aligned}
\label{eq:BWT_eq}
\end{equation}

Now we consider the write count of the weight matrix $W$ in the $j^\mathrm{th}$ layer, which is partitioned into a set $B$ with $P\times Q$ blocks of size $k\times k$.
Each PTC $t$ in a cluster $C$ is assigned with $n_t$ blocks, i.e., $\{B_{1}^{t}, B_{2}^{t}, \dots, B_{n_t}^{t}\}$.
The number of writes for layer $j$ (LWT) is computed as
\begin{equation}
\small
LWT^{j} = \sum_{t}^{C}\sum_{i}^{n_t} WT(B_{i}^{t}, B_{i-1}^{t}),
\label{eq:LWT_eq}
\end{equation}
where we assume the first block is mapped onto the initialized PTC with all PCM wires being set to $c$ state.

Hence, to deploy a model, the total number of writes is the sum of layer-wise write operations, i.e., $\#total\ writes = \sum_j^L LWT^{j}$.
To evaluate the status of the most over-utilized memory cell, we define a layer-wise metric, \emph{\# max writes}, by counting the maximum number of write operations over a single PCM memory cell for a cluster of PTCs.
Table.~\ref{tab:VGG85bitLayerData} shows the statistics of the number of writes and the maximal writes for the convolutional layers of 5-bit VGG8.
Though our augmented redundant write elimination strategy is applied, we still observe a significant number of writes that challenge the PCM endurance.
Therefore, in the following discussion, several techniques are proposed to minimize both \emph{\# total writes} and \emph{\# max writes}.
\begin{figure}
    \centering
    \includegraphics[width=0.48\textwidth]{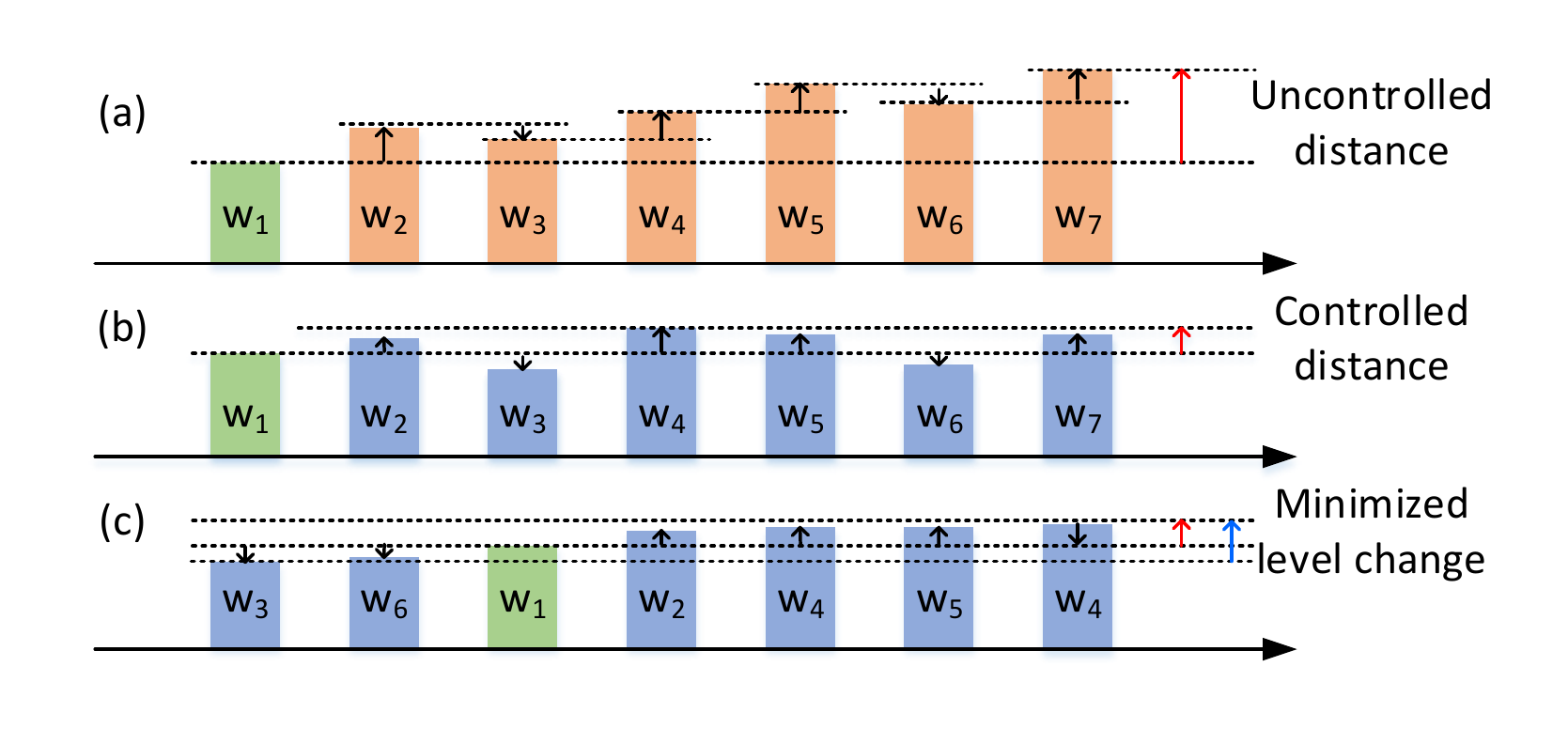}
    \caption{~\small An example of constraining a weight sequence starting from $w_{1}$.
    (a) Constrain the distance between neighbors.
    (b) Constrain weights around reference value ($w_1$).
    (c) Sort weights in (b) in an ascending order.}
    \label{fig:DiffLossScheme}
\end{figure}

\begin{figure}
    \centering
    \includegraphics[width=0.5\textwidth]{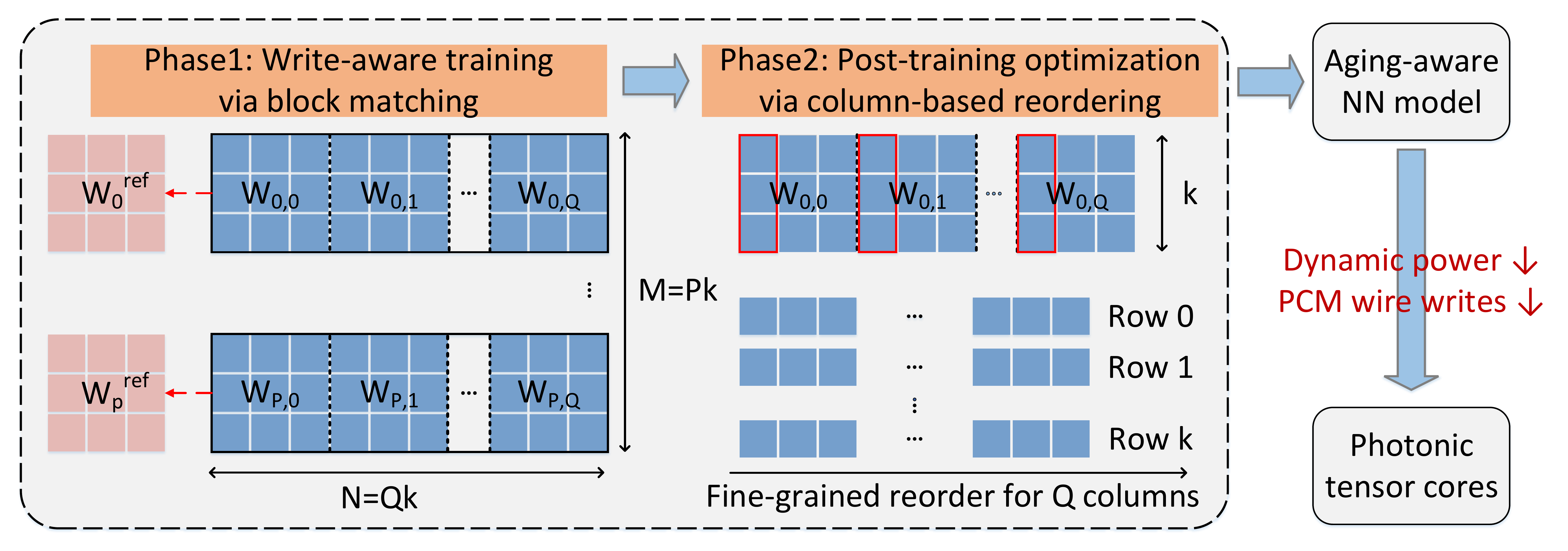}
    \caption{
    ~\small Proposed two-phase \texttt{ELight} to enable efficient photonic in-memory neurocomputing with lifetime enhancement.}
    \label{fig:ElightFlow}
\end{figure}

\begin{table}[]
\centering
\caption{\small Layer-wise statistics of writes of 5-bit VGG8 model.}
\label{tab:VGG85bitLayerData}
\resizebox{0.45\textwidth}{!}{
\begin{tabular}{|c|c|c|c|c|}
\hline
 & Layer 2 & Layer 3 & Layer 4 & Layer 5 \\ \hline
\emph{\# total writes} & 1.14$\times 10^{6}$ & 4.87$\times 10^{6}$ & 1.66$\times 10^{7}$ & 3.26$\times 10^{7}$ \\ \hline
\emph{\# max writes} & 294 & 534 & 914 & 1425 \\ \hline
\end{tabular}
}
\end{table}

\subsection{Write-Aware Training via Block Matching}
To reduce the number of write operations when mapping a group of weight blocks, 
a straightforward technique is to shrink the weight distance between neighboring blocks.
Take the example of reducing write operations when writing a weight sequence into one memory cell.
As shown in Fig.~\ref{fig:DiffLossScheme}(a), by constraining the distance between neighbors, neighboring weights are encouraged to be similar, thus holding more redundant writes.
However, it sets no constraint on the distance between the largest value and the smallest value.
This requires more wires to be programmed to demonstrate a wider range.
Hence, a wise solution is to constrain weights around a reference value such that the value range is under control, shown in Fig.~\ref{fig:DiffLossScheme}(b).

To boost the similarity among weight blocks, we propose a write-aware training procedure, shown as phase 1 in Fig.~\ref{fig:ElightFlow}.
We set the average block as a reference for weight blocks assigned to the same PTC and penalize their transmission level distance from the reference.
Instead of directly optimizing
the level difference between blocks using \eqref{eq:BWT_eq} in a L1 way, we use an L2 regularization term to calculate the level difference (LD) between block $B$ and $A$ as, 
\begin{equation}
\small
\begin{aligned}
    LD(B, A)\!=\!\sum_{i}^{k}\!\sum_{j}^{k}\!\lVert \tilde{l}^{+}(b_{ij})\!-\!\tilde{l}^{+}(a_{ij})\rVert^2\!+\!\lVert \tilde{l}^{-}(b_{ij})\!-\! \tilde{l}^{-}(a_{ij})\rVert^2.
\end{aligned}
\end{equation}
\begin{equation}
\small
\begin{aligned}
    LD(W^{ref}, W)\!=\!\sum_{i}^{k}\!\sum_{j}^{k}\!\lVert \tilde{l}^{+}(w_{ij}^{ref})\!-\!\tilde{l}^{+}(w_{ij})\rVert^2\!+\!\lVert \tilde{l}^{-}(w_{ij}^{ref})\!-\! \tilde{l}^{-}(w_{ij})\rVert^2.
\end{aligned}
\end{equation}
Here, $\tilde{l}^{+}$ and $\tilde{l}^{-}$ denote the normalized $l^{+}$ and $l^{-}$ by dividing $\alpha_b = 2^b -1$.
Normalizing level data to $[-1, 1]$ can help healthy gradient propagation.
The L2 regularization term imposes a heavier penalty for large value deviation, and slight deviation is allowed to maintain diverse weights.
In this way, not only \emph{\# max writes} can be constrained via rejecting large value deviation,
but the model expressivity can be mostly maintained.
Thus, the block matching loss is defined as
\begin{equation}
\small
\begin{aligned}
    \mathcal{L}_{BM} &= \sum_{l}^{L}\sum_{t}^{G}\sum_{i}^{n_g} \frac{1}{\beta^{B}} LD(B_{i}^t, B^t_{avr}),
\end{aligned}
\end{equation}
where we explicitly encourage the similarity of weight blocks in the same group $G$. 
$\beta^B$ is the block size to normalize the level distance.

By adding the block matching loss to the loss function, we trade off between the accuracy and block similarity by controlling $\lambda$, as, 
\begin{equation}
\small
\begin{aligned}
    \mathcal{L} &= \mathcal{L}_{CE} + \lambda \mathcal{L}_{BM},
\end{aligned}
\end{equation}
where $\mathcal{L}_{CE}$ is the original cross-entropy loss.

However, there are three issues to optimize $\mathcal{L}_{BM}$.
First, it is not differentiable due to the \texttt{Round}($\cdot$) operations in \eqref{eq:level_pos} and \eqref{eq:level_neg}.
Second, the gradient approximation through the \emph{logarithmic} operations needs to be carefully handled.
Third, transmission levels in both positive and negative PTCs are used to compute $LD$, while only levels that are physically implemented on PTCs need to be involved in gradient evaluation.
In other words, only the gradient from either $\lVert \tilde{l}^{+}(w) - \tilde{l}^{+}(\delta)\rVert^2$ or $\lVert \tilde{l}^{-}(w) - \tilde{l}^{-}(\delta)\rVert^2$ need to be propagated back to compute the gradient w.r.t $w$ depending on its sign.
Hence, by leveraging straight-through-estimator (STE) to approximate the gradients for $\texttt{Round}(\cdot)$, we propagate the gradient of $\mathcal{L}_{BM}$ back as,
{\small
\begin{align}
\begin{aligned}
    \frac{\partial \mathcal{L}_{BM}}{\partial W}\!\!=\!\!\frac{1}{\beta^B}\!\!(\frac{\partial \mathcal{L}_{BM}}{\partial \tilde{l}^{+}(W)}\!\frac{\mathrm{d} \tilde{l}^{+}(W)}{\mathrm{d} W}\!\odot\! \mathcal{M}^{+}\!+\!
    \frac{\partial \mathcal{L}_{BM}}{\partial \tilde{l}^{-}(W)}\! \frac{\mathrm{d} \tilde{l}^{-}(W)}{\mathrm{d} W}\!\odot\!\! \mathcal{M}^{-}\!\!),
\end{aligned}
\end{align}
}
where 
$\mathcal{M}^{+}$ and $\mathcal{M}^{-}$ are non-negative and negative masks of $W$ to extract the needed gradients from $\frac{\mathrm{d} \tilde{l}^{+}(W)}{\mathrm{d} W}$ and $\frac{\mathrm{d} \tilde{l}^{-}(W)}{\mathrm{d} W}$, computed by,
{\small
\begin{align}
    &\begin{aligned}
    &\left.\frac{\mathrm{d} \tilde{l}^{+}(W)}{\mathrm{d} W} \right\vert _{W\geq0}
    = \frac{-\mathrm{d}(\log_t(s|W|+\delta))}{\alpha_{b}\mathrm{d} W} = \frac{-s}{\alpha_{b}\ln(t)(s|W|+\delta)} \\
    \end{aligned}, \\
    &\begin{aligned}
    &\left.\frac{\mathrm{d} \tilde{l}^{-}(W)}{\mathrm{d} W} \right\vert _{W<0}
    = \frac{\mathrm{d}(\log_t(s|W|+\delta))}{\alpha_{b}\mathrm{d} W} = \frac{-s}{\alpha_{b}\ln(t)(s|W|+\delta)} \\
    \end{aligned}.
\end{align}
}

\subsection{Post-Training Optimization via Column-based Reordering}
While the technique proposed above boosts the similarity among weight blocks, 
there is still room for further optimization since it does not consider the mapping order of blocks.
We still take Fig.~\ref{fig:DiffLossScheme}(b) as an example.
The sum of neighboring differences is not explicitly optimized as we only limit all weights around one reference value.
One straightforward method to further reduce the sum of neighboring differences is to sort the weight sequence either in ascending or descending order, as shown in Fig.~\ref{fig:DiffLossScheme}(c).

Inspired by this heuristic, we propose a fine-grained \emph{column-based reordering} method to sort the weights in blocks that share the same PCM memory cells in PTCs.
The second phase in Fig.~\ref{fig:ElightFlow} illustrates our idea:
For a group of weight blocks assigned to one PTC,
weights located in the same position are first shaped into 1-D sequences, then weight sequences are separately sorted in ascending order.
We can also sort them in descending order.
The weights are finally scattered back to different blocks in the new order.
Note that the above process is equivalent to swapping columns element-wise with no influence on final results.
Besides, with the aid of the sorting heuristic, when mapping a group of blocks, weight values are written into photonic memories in ascending order.
By eliminating redundant writes over wires, \emph{\# max writes} over single photonic memory cell can be upper-bounded by the level range of its stored weights, i.e., $2^{b+1}$$-$$1$.

Therefore, combining write-aware training with post-training optimization, \emph{\# total writes} and \emph{\# max writes} can be significantly reduced to mitigate the aging issue and the tedious programming efforts.

\section{Experimental Results}
\label{sec:ExperimentalResults}

To demonstrate the effectiveness of proposed two-phase framework \name,  we conduct experiments on MNIST~\cite{NN_MNIST1998}, FashionMNIST~\cite{NN_FashionMNIST2017}, CIFAR-10 and CIFAR-100~\cite{NN_cifar2009} datasets.
On the first two tasks, a CNN configuration C32K4-C32K4-P5-F64-F10
is adopted, where C32K4 is a $4\times 4$ convolutional layer with 32 kernels, P5 means average pooling with output size $5\times 5$, and F64 is a fully-connected (FC) layer with 64 neurons.
On CIFAR-10 and CIFAR-100, VGG8~\cite{NN_NN2018_Dong}
and VGG13~\cite{NN_Arxiv2014_Simonyan} are used, while we replace the last three FC layers with one FC layer to avoid over-fitting.
We implement all models in PyTorch on a machine with an Intel Core i7-9700 CPU and an NVIDIA Quadro RTX 6000 GPU.
The CNN and VGG models are trained for 100 and 200 epochs, respectively, using the SGD optimizer with a momentum of 0.9.
Regarding the photonic tensor core size, we assume $16\times16$ and $64\times64$ for the small CNN and VGG model, respectively.
Note that for fair comparison, the \emph{augmented redundant write elimination} (ARWE) strategy is also applied in the baseline to collect statistics.

\subsection{Evaluation of the Distribution-Aware Quantization Scheme}
\begin{figure}
    \centering
    \vspace{-5pt}
    \subfloat[]{\includegraphics[width=0.235\textwidth]{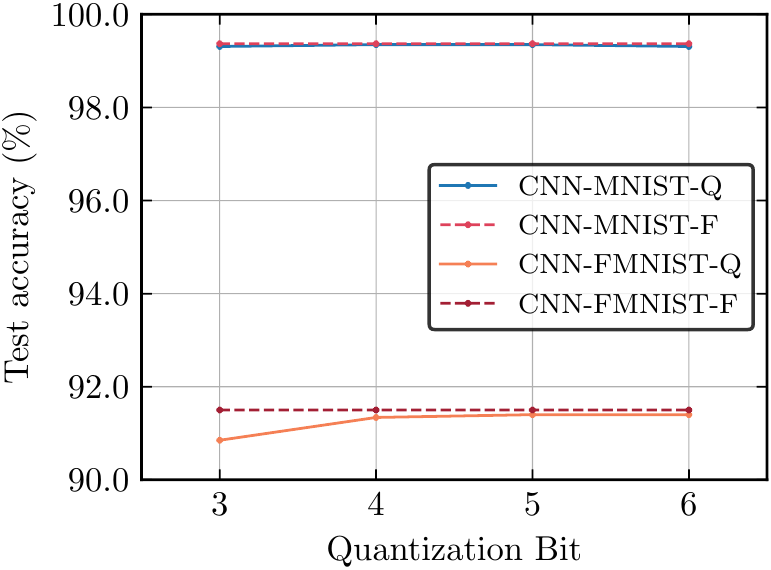}
    \label{fig:QuantAccCNN}
    }
    \subfloat[]{\includegraphics[width=0.235\textwidth]{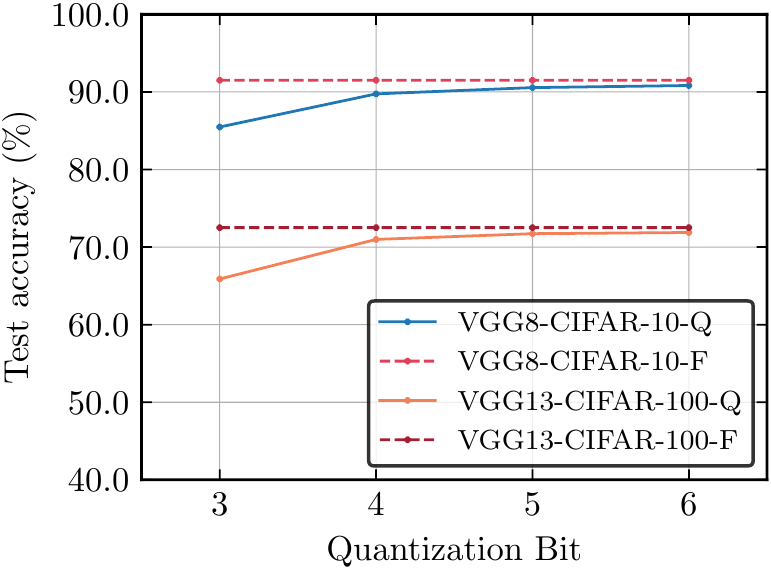}
    \label{fig:QuantAccVGG}
    }
    \caption{~\small 
   Quantization evaluation on (a) CNN and (b) VGG models. F means full-precision models. Q means models with quantization.
   }
    \label{fig:QuantAcc}
    \vspace{-10pt}
\end{figure}

In Fig.~\ref{fig:QuantAcc}, we evaluate our augmented base-$c$ quantizer under 3- to 6-bit quantization.
Our quantization scheme can successfully tackle the issue of non-linear transmission level distribution, enlarges the solution space, and achieves high accuracy under low-bit quantization.
Under 4- to 6-bit quantization, our proposed method can achieve negligible or small accuracy loss on all tasks.
Our method can still maintain $>85\%$ on CIFAR-10 and $>65\%$ accuracy on CIFAR-100 under 3-bit quantization for relatively complicated tasks.

\begin{table*}[]
\centering
\caption{\small Performance of ELight on VGG networks on CIFAR-10 and CIFAR-100 dataset. AC: accuracy change, R: column-based reordering. The \emph{\# max writes} of one largest convolutional layer (the 5$^\mathrm{th}$ convolutional layer for VGG8 and 8$^\mathrm{th}$ convolutional layer for VGG13) is shown here.}
\label{tab:resElight}
\resizebox{0.75\textwidth}{!}{%
\begin{tabular}{|c|c|c|c|c|c|c|c|c|c|c|}
\hline
\multirow{2}{*}{Network} & \multirow{2}{*}{Dataset} & \multirow{2}{*}{Bitwidth} & \multirow{2}{*}{$\lambda$} & \multirow{2}{*}{Acc($\%$)/AC} & \multicolumn{2}{c|}{\emph{\# total writes} $\downarrow (\times)$}  & \multicolumn{2}{c|}{Energy cost $\downarrow (\times)$} & \multicolumn{2}{c|}{\emph{\# max writes}}\\ \cline{6-11} 
 &  &  &  &  & - & +R & - & +R & - & +R \\ \hline\hline
\multirow{8}{*}{VGG8} & \multirow{8}{*}{CIFAR-10} & \multirow{2}{*}{3} & 0 & 86.71 & 1 & 6.52 & 1 & 9.27 & 128 & 15 \\ \cline{4-11} 
 &  &  & 8 & 86.02/-0.69 & 22.12 & \textbf{46.11} & 6.63 & \textbf{69.29} & 14 & \textbf{7} \\ \cline{3-11} 
 &  & \multirow{2}{*}{4} & 0 & 89.75 & 1 & 7.84 & 1 & 11.31 & 401 & 36 \\ \cline{4-11} 
 &  &  & 10 & 89.94/+0.19 & 3.83 & \textbf{24.45} & 3.92 & \textbf{35.48} & 95 & \textbf{19} \\ \cline{3-11} 
 &  & \multirow{2}{*}{5} & 0 & 90.56 & 1 & 10.01 & 1 & 14.35 & 1425 & 82 \\ \cline{4-11} 
 &  &  & 10 & 90.12/-0.44 & 3.17 & \textbf{22.28} & 3.20 & \textbf{31.17} & 494 & \textbf{74} \\ \cline{3-11} 
 &  & \multirow{2}{*}{6} & 0 & 90.83 & 1 & 12.31 & 1 & 16.89 & 4464 & 180 \\ \cline{4-11} 
 &  &  & 5 & 89.88/-0.95 & 6.82 & \textbf{26.35} & 7.15 & \textbf{32.48}  & 1560 & \textbf{146} \\ \hline\hline
\multirow{6}{*}{VGG13} & \multirow{6}{*}{CIFAR-100} & \multirow{2}{*}{4} & 0 & 70.99 & 1 & 9.66 & 1 & 13.84  & 542 & 39\\ \cline{4-11} 
 &  &  & 10 & 70.44/-0.55 & 3.54 & \textbf{29.25} & 3.57 & \textbf{42.02} & 173 & \textbf{33} \\ \cline{3-11} 
 &  & \multirow{2}{*}{5} & 0 & 71.73 & 1 & 12.06 & 1 & 17.29 & 1771 & 84\\ \cline{4-11} 
 &  &  & 3 & 71.95/+0.22 & 2.19 & \textbf{21.93} & 2.21 & \textbf{31.41} & 921 & \textbf{55} \\ \cline{3-11} 
 &  & \multirow{2}{*}{6} & 0 & 71.88 & 1 & 14.37 & 1 & 17.62  & 4926 & 182 \\ \cline{4-11} 
 &  &  & 3 & 70.97/-0.91 & 3.11 & \textbf{22.65} & 3.19 & \textbf{29.85} & 3577 & \textbf{156} \\ \hline
\end{tabular}}
\end{table*}

\subsection{Evaluation of Proposed Aging-Aware Optimization Framework}
\subsubsection{Evaluation of write-aware training via block matching}
\begin{figure}
    \centering
    \vspace{-5pt}
    \subfloat[]{\includegraphics[width=0.245\textwidth]{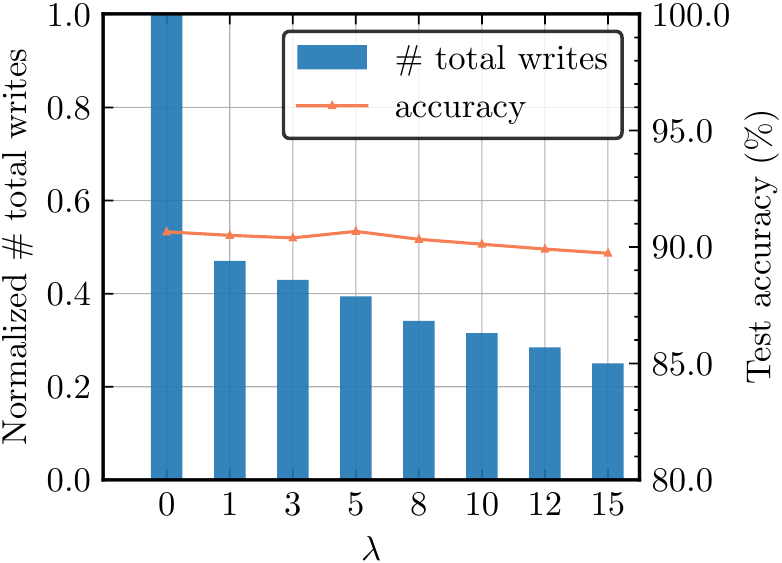}
    \label{fig:BlockMatchScanVGG84bit}
    }
    \subfloat[]{\includegraphics[width=0.245\textwidth]{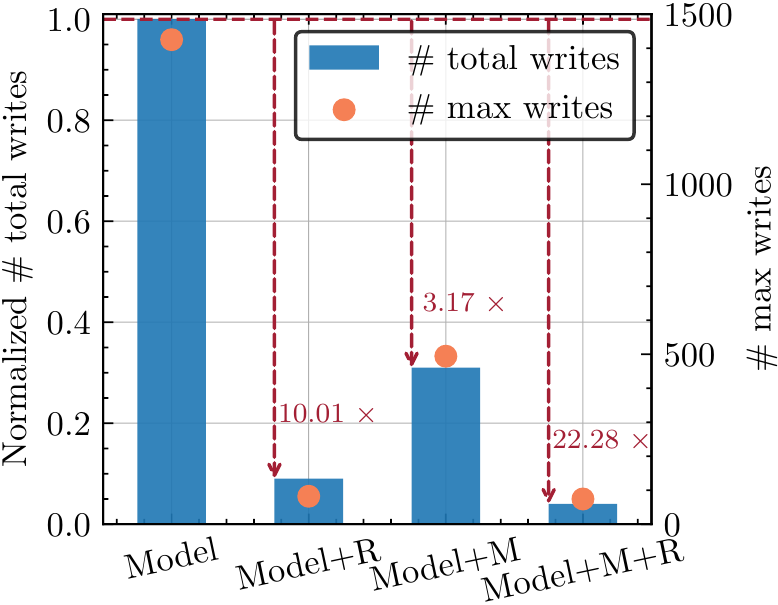}
    \label{fig:ColumnReorderVGG85bit}
    }
    \caption{~\small
    Evaluation of proposed aging-aware optimization techniques on 5-bit VGG8.
   (a) Normalized \emph{\# total writes} and accuracy comparison with different $\lambda$ for write-aware training.
   (b) Comparison between \emph{\# total writes} and \emph{\# max writes} of the 5$^\mathrm{th}$ convolutional layer.
   }
    \label{fig:EvaluationTech}
\end{figure}

To investigate the impact of write-aware training, 
we visualize the normalized \emph{\# total writes} and accuracy of 5-bit VGG8 with various degrees of $\lambda$ in Fig.~\ref{fig:BlockMatchScanVGG84bit}.
With the increase of $\lambda$, \emph{\# total writes} decreases since of the existence of stronger similarity among blocks.
With too large $\lambda$, accuracy starts to decline since weights cannot be identical even we pursue block-level similarity.
Otherwise, it will be hard for weights to capture diverse features.
By trading off the accuracy and \emph{\# total writes}, a sweet point ($\lambda=10$) exists with 3.17$\times$ reduction in \emph{\# total writes} and only $0.44\%$ accuracy drop.
Interestingly, sometimes a proper $\lambda$ leads to better accuracy as a regularization mechanism.

\subsubsection{Evaluation of post-training optimization via column-based reordering}
Fig.~\ref{fig:ColumnReorderVGG85bit} shows the comparison of \emph{\# total writes} of 5-bit VGG8 between (1) Model trained with/without column-based reordering (R), (2) Model trained with block matching loss (M) with/without column-based reordering (R). 
We also compare the \emph{\# max writes} of mapping the $5$-th convolutional layer.
The results provide three insights:
(1) The column-based reordering can reduce \emph{\# total writes} a lot even without extra training efforts to boost weight similarity, achieving $10.01\times$ reduction.
(2) With the write-aware training to boost similarity among blocks, the column-based reordering can achieve the best of reduction on \emph{\# total writes} by $22.28\times$, which justifies the effectiveness of our joint aging-aware optimization framework.
(3) Our proposed column-based reordering upper-bounds \emph{\# max writes} by the number of transmission levels, i.e., $2^{b+1}$$-$$1$, which eliminates any redundant writes and ensures that only a small number of PCM wires need to be re-written.

\subsubsection{Evaluation on the synergistic optimization framework}
\begin{figure}
    \centering
    \vspace{-5pt}
    \subfloat[]{\includegraphics[width=0.245\textwidth]{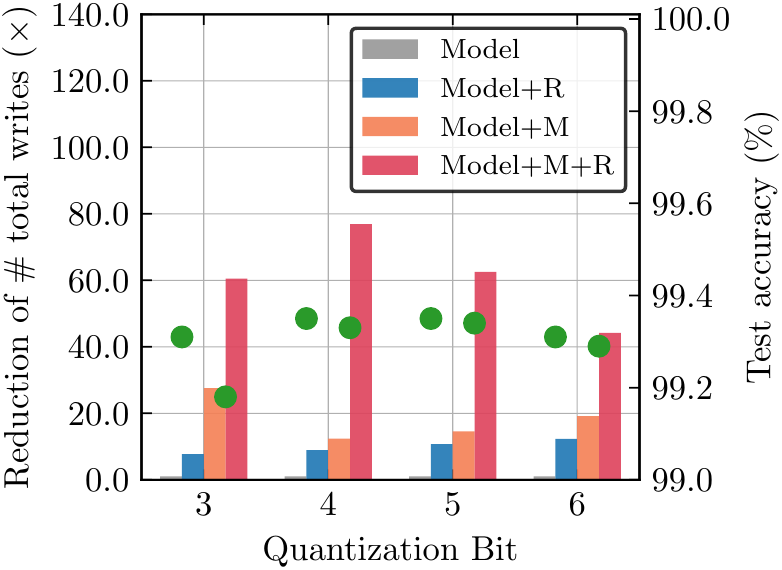}
    \label{fig:/WTReductionMNIST}
    }
    \subfloat[]{\includegraphics[width=0.245\textwidth]{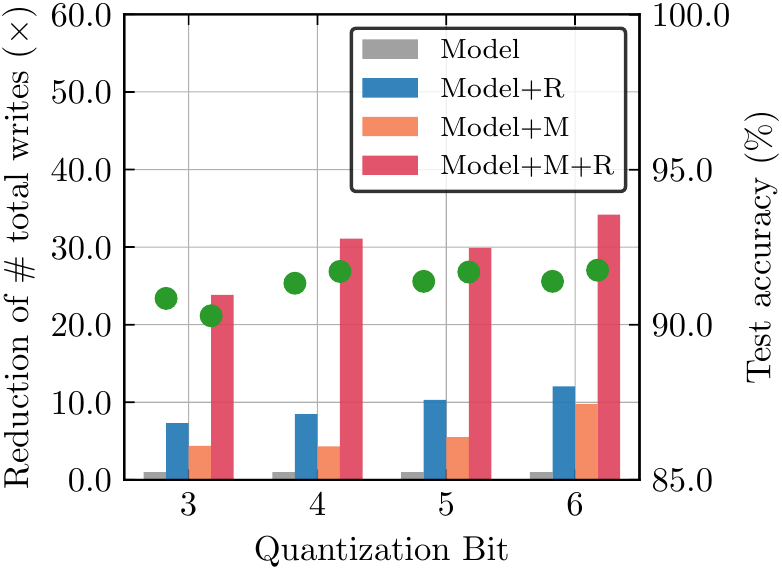}
    \label{fig:/WTReductionFMNIST}
    }
    \caption{~\small 
   The reduction of \# total writes and accuracy of CNN model trained on (a) MNIST and (b) FashionMNIST under different bit-width.}
    \label{fig:WTReductionCNN}
    \vspace{-10pt}
\end{figure}

To further testify the effectiveness of the proposed aging-aware optimization framework, we evaluate different models under different bit-width quantization.
For practical use, we choose a sweet point of $\lambda$
to constrain the accuracy drop within $1\%$.
Fig.~\ref{fig:WTReductionCNN} shows the reduction of \emph{\# total writes} and accuracy of the CNN model on MNIST and FashionMNIST under 3- to 6-bit quantization.
On those simple tasks, a significant reduction of \emph{\# total writes} is obtained with negligible accuracy drop.
Table.~\ref{tab:resElight} shows the effectiveness of our proposed techniques on VGG8 and VGG13 trained on CIFAR-10 and CIFAR-100, respectively.
The proposed write-aware training and post-training optimization
techniques can work orthogonally to each other, achieving the largest reduction on \emph{\# total writes} and \emph{\# max writes}, where \emph{\# total writes} can be reduced by $>20\times$ with less than $1\%$ accuracy degradation.
Hence, our joint optimization framework can successfully mitigate the aging issue by largely reducing the number of write operations.

\begin{table}[]
\centering
\caption{\small Pulse profiles for $a\!\rightarrow\!c$ and $c\!\rightarrow\!a$ transition~\cite{NP_APR2020_Miscuglio}.}
\label{tab:progPulses}
\resizebox{0.4\textwidth}{!}{
\begin{tabular}{|c|c|c|c|}
\hline
 & Pulse period($\mu s$) & Pulse voltage($V$) & \# Pulse  \\ \hline
$a\!\rightarrow\!c$ & 1 & 5 & 20  \\ \hline
$c\!\rightarrow\!a$ & 0.5 & 15 & 1\\
\hline
\end{tabular}
}
\end{table}

\subsubsection{Evaluation of power saving}
To testify the energy efficiency when applying the above optimization methods, we trace the detailed energy cost of writing weight block data onto PTCs during the inference process.
Assuming the resistance of heaters is consistent, the ratio of write energy cost between a-c and c-a transition is $40:9$ using the pulse profiles on external heaters in Table~\ref{tab:progPulses}.

Thanks to the reduction of \emph{\# total writes}, the energy cost of deploying VGG8 and VGG13 is reduced by $> 30\times$ under different bit-widths, as shown in Table.~\ref{tab:resElight}, which proves our aging-aware optimization framework effectively saves dynamic energy cost incurred by PTC reuse.

\section{Conclusion}
\label{sec:Conclusion}
In this work, we propose a synergistic solution to enable efficient photonic in-memory neurocomputing with an enhanced lifetime.
First, we model the non-linear transmission distribution of PCM-based photonic memories and propose a distribution-aware quantization scheme to reduce weight encoding errors.
Based on this, we propose a synergistic aging-aware optimization framework \name to trim down redundant PCM writes via block-matching-based training and column-based reordering.
Our co-optimization method significantly reduces the number of total write operations and the number of write operations for the most over-utilized memory cell. 
Experimental results demonstrate that the proposed optimization framework can reduce the number of write operations and energy costs by $>$$20\times$,
pushing photonic in-memory neurocomputing towards practical application in efficient machine learning.
\section*{Acknowledgement}
The authors acknowledge the Multidisciplinary University Research Initiative (MURI) program through the Air Force Office of Scientific Research (AFOSR), contract No. FA 9550-17-1-0071, monitored by Dr. Gernot S. Pomrenke.

{\small
\balance
\bibliographystyle{IEEEtran}
\bibliography{./ref/Top_sim,./ref/NN,./ref/NP,./ref/ALG, ./ref/addition,./ref/IEEEsettings}
}

\end{document}